\documentstyle[epsfig,aps,twocolumn]{revtex}

\newcommand{\pol}{\hat{\bf e}}
\newcommand{\rv}{{\bf r}}
\newcommand{\Ev}{{\bf E}}
\newcommand{\Dv}{{\bf D}}
\newcommand{\Pv}{{\bf P}}

\newcommand{\dv}{{\bf d}}

\newcommand{\Dkappav}{\Delta {\bbox \kappa}}

\newcommand{\kv}{{\bf k}}

\newcommand{\eo}{\epsilon_0}
\newcommand{\beq}{\begin{equation}}
\newcommand{\eeq}{\end{equation}}
\newcommand{\bea}{\begin{eqnarray}}
\newcommand{\eea}{\end{eqnarray}}

\renewcommand{\>}{\rangle}
\renewcommand{\(}{\left(}
\renewcommand{\)}{\right)}
\renewcommand{\[}{\left[}
\renewcommand{\]}{\right]}

\newcommand{\commentout}[1]{{}}

\begin{document}

\draft
\preprint{}
\title{Optical linewidth of a low density Fermi-Dirac gas}
\author{Janne Ruostekoski$^1$ and Juha Javanainen$^2$}
\address{$^1$Abteilung f\"ur Quantenphysik,
Universit\"at Ulm, D-89069 Ulm, Germany\\
$^2$Department of Physics, University of Connecticut, Storrs,
Connecticut 06269-3046}
\date{\today}
\maketitle
\begin{abstract}
We study propagation of light in a Fermi-Dirac gas at zero
temperature. We analytically obtain
the leading density correction to the optical linewidth.
This correction is a direct consequence of the quantum statistical
correlations of atomic positions that modify the optical
interactions between the atoms at small interatomic separations.
The gas exhibits a dramatic line narrowing already at very low densities.

\end{abstract}

The observation of Bose-Einstein condensation in dilute atomic vapors
\cite{becexp} has spurred much interest in ultracold atomic gases. Another
evident milestone of atomic physics would be the cooling of a Fermi-Dirac
(FD) gas to the quantum degenerate regime. So far all probing of atomic
Bose-Einstein condensates has been done optically, and obviously optical
detection could also play an important role in the experimental studies
of FD  gases. Appropriately, theoretical studies of the FD gases are
experiencing a renaissance~\cite{JAV95b,STO96,BAR96,MAR98,JAV99}.

In this paper we investigate propagation of low-intensity light in a FD
gas in the limit of low atom density. We derive the leading quantum
statistical correction to the standard column-density refractive index
analytically, by legitimately ignoring collective linewidths and line
shifts generated in the processes in which a photon is repeatedly
scattered between the same atoms. A fermion gas exhibits a striking line
narrowing characteristic of the FD statistics, which behaves as
$\rho^{2/3}$ at low atom densities.

In the dipole approximation it is advantageous to transform the
Hamiltonian into the {\it length\/} gauge by the  Power-Zienau-Woolley
transformation~\cite{COH89}. Then the positive frequency component of the
electric field $\Ev^+$ may be expressed~\cite{RUO97a,RUO97c} in terms of
the positive frequency components of the driving electric displacement,
${\bf D}^+_F$, and of the source field radiated by atomic polarization,
${\bf P}^+$, as
\begin{mathletters}
\bea
\eo{\bf E}^+({\bf r})& =& {\bf D}^+_F({\bf r}) +
{1\over i\kappa}\int d^3r'\,
{\sf G}({\bf r}-{\bf r'})\,{\bf P}^+({\bf r}')\,,
\label{eq:MonoD}\\
{\sf G}_{ij}({\bf r})& =& i\kappa\left\{
\left[ {\partial\over\partial r_i}{\partial\over\partial r_j} -
\delta_{ij} {\bbox \nabla}^2\right] {e^{ikr}\over4\pi r}
-\delta_{ij}\delta({\bf r})
\right\}\,.
\label{eq:GDF}
\eea
\label{eq:atomlight}
\end{mathletters}
Here $k = {\Omega / c}$, $\Omega$ is the frequency of the driving field,
and the scalar constant $\kappa =
{\cal D}^2/\hbar\epsilon_0$ is defined in terms of the reduced dipole
moment matrix element ${\cal D}$. The monochromatic dipole radiation
kernel
${\sf G}(\rv)$ coincides with the corresponding classical
expression \cite{JAC75}. In second quantization the polarization reads
\begin{equation}
{\bf P}^+({\bf r}) = {\bf d}_{ge}\psi^\dagger_{g}({\bf
r})\psi_{e}({\bf r})\,.
\label{eq:PDF}
\end{equation}
Here $\psi_g$ and $\psi_e$ are the ground state and the excited state atom
field operators in the Heisenberg picture, and $\dv_{ge}$ is the dipole
matrix element for the transition $g\rightarrow e$.
For simplicity, we consider here two-level atoms with just a single
ground state $|g\>$ and one excited state $|e\>$, using a constant real
vector $\dv$ (such that ${\cal D}=|\dv|$) as the dipole matrix element.

While Eqs.~{(\ref{eq:atomlight})} describe the scattered light in a
medium, in general, with a small atom-light detuning and for a dense
atomic sample, there is no easy way to find the polarization
$\Pv^+(\rv)$. By making a field theory version of the Born and
Markov approximations, we have
derived a hierarchy of equations of motion for correlation functions
that contain one excited-atom field and one, three, five, etc., ground
state atom fields, for the limit of low light intensity~\cite{RUO97a}.
In the present case of two-level atoms the hierarchy reads
\bea
\lefteqn{\dot{\bf P}_l({\bf r}_1,\cdots,\rv_{l-1};{\bf r}_l)=
(i\delta-\gamma) {\bf P}_l({\bf r}_1,\cdots,\rv_{l-1};{\bf
r}_l)}\nonumber\\
&&+\sum_{k=1}^{l-1}{\sf P}\cdot{\sf G}'({\bf r}_l-{\bf r}_k) {\bf
P}_l({\bf r}_1,\cdots,\rv_{k-1},\rv_{k+1},\cdots,\rv_l;{\bf r}_k)
\nonumber\\
&&+i\kappa \rho_l({\bf r}_1,\cdots,{\bf r}_l) {\sf P}\cdot
{\bf D}^+_F({\bf r}_l)\nonumber\\
&&+\int d^3r_{l+1}\,{\sf P}\cdot{\sf G}'({\bf r}_l-
{\bf r}_{l+1})
{\bf P}_{l+1}({\bf r}_1,\cdots,{\bf r}_l;{\bf r}_{l+1})\,,
\label{eq:hie}
\eea
where $\gamma={\cal D}^2 k^3/(6\pi\hbar\eo)$ denotes the spontaneous
linewidth, and $\delta$ is the atom-light detuning. We have defined a
projection operator
\beq
{\sf P}\equiv {{\bf d}{\bf d}\over|\dv|^2}\,,
\eeq
whose purpose is to eliminate all but the two atomic states from
consideration, and the correlation functions
\begin{mathletters}
\bea
\lefteqn{{\bf P}_l({\bf r}_1,\cdots,{\bf r}_{l-1};{\bf r}_l)}
\nonumber\\&&\equiv
\langle \psi^\dagger_g({\bf r}_1)\cdots\psi^\dagger_g({\bf r}_{l-1})
{\bf P}^+({\bf r}_l) \psi_g({\bf r}_{l-1})\cdots
\psi_g({\bf r}_1)\rangle,\\&&
\rho_l({\bf r}_1,\cdots,{\bf r}_l) \equiv \langle
\psi_g^\dagger({\bf r}_1)\cdots
\psi_g^\dagger({\bf r}_l) \psi_g ({\bf r}_l)\cdots \psi_g ({\bf r}_1)
\rangle.
\eea
\label{eq:expect}
\end{mathletters}
The quantity ${\bf P}_l$ reflects correlations between the dipole moment
of one
atom and the positions of $l-1$ other atoms, and $\rho_l$ is simply the
density correlation function for $l$ ground state atoms.

The terms in the sum on the right-hand side of Eqs.~{(\ref{eq:hie})}
represent processes in which the $l$ atoms at $\rv_1,\ldots,\rv_l$ repeatedly
exchange photons. Such processes are the microscopic mechanism for
collective linewidths and line shifts. The integral stands for a
process in which yet another atom shines its light on the atom at $\rv_l$.

Due to the resulting divergent dipole-dipole interactions, all
correlation functions $\Pv_l$ vanish whenever two position arguments are
the  same~\cite{RUO97c}.  The Lorentz-Lorenz local-field correction
follows mathematically from this observation. Moreover, without changing
the outcome of the hierarchy, we may, and will, remove  all  contact
interactions between different atoms in Eqs.~{(\ref{eq:hie})} by
introducing the field  propagator ${\sf G}'$ defined by
\begin{equation}
{\sf G}_{ij}'(\rv)={\sf G}_{ij}(\rv)+i\kappa\delta_{ij}\delta(\rv)/3\,.
\end{equation}
This definition indicates that the integral of ${\sf G}'$ over an
infinitesimal volume enclosing the origin vanishes.

The coupled theory for light and matter fields
[Eqs.~{(\ref{eq:atomlight})} and {(\ref{eq:hie})}] may be solved, in
principle exactly, by means of stochastic simulations \cite{JAV99}.
This is because the correlation hierarchy~(\ref{eq:hie}) is the same as
the hierarchy describing the classical electrodynamics of charged
harmonic oscillators with the position correlations $\rho_l$. By
synthesizing a stochastic ensemble of samples of dipoles that have the
position correlation functions $\rho_l$ and calculating the
ensemble-averaged response to classical light, we have a solution to
Eqs.~(\ref{eq:hie}). Unfortunately, such simulations are demanding on
computer time. The computations of Ref.~\cite{JAV99} were therefore
performed within a one-dimensional (1D) model  electrodynamics. While the
predictive power of 1D electrodynamics may be questioned, the simulation
results for a FD gas at
$T=0$ show clear signatures of the quantum statistics:  Even in the
limit of zero density, the optical linewidth of the FD gas is only half of
the resonance linewidth of an isolated atom~\cite{JAV99}.

The 1D simulations have also allowed us to test predictions of
the density expansion introduced by Morice {\it et al.}~\cite{MOR95} in
their studies of the optical response of a quantum degenerate
Bose-Einstein gas. At least in one dimension this expansion is in
semi-quantitative agreement with numerical simulations, and in the
low-density limit the agreement is excellent~\cite{JAV99}.
With this in mind, we venture to use the approximation of Morice {\it et
al.}~\cite{MOR95} to truncate the correlation hierarchy {(\ref{eq:hie})}
also in the present three-dimensional case.

We consider the steady-state solution of {(\ref{eq:hie})}. The atoms
are assumed to fill the half-infinite space $z>0$ with a constant
density $\rho$. The incoming free field is written $\Dv_F(\rv)=
D_F \,\pol \exp{(ikz)}$, and we assume that $\pol || {\bf d}$.
The hierarchy of equations {(\ref{eq:hie})} is
truncated by writing~\cite{MOR95}
\beq
\Pv_3(\rv_1,\rv_2;\rv_3)\simeq {\rho_2(\rv_1,\rv_2)\over \rho}
\Pv_2(\rv_2;\rv_3)\,.
\label{appr}
\eeq
The pair correlation function is written in the form
\begin{equation}
\rho_2({\bf r}_1,{\bf r}_2) = \rho^2\,[1+\varphi({\bf r}_1-{\bf
r}_2)]\,.
\label{eq:DCF}
\end{equation}
By introducing the dimensionless quantities
\beq
\bar\delta\equiv\delta/\gamma,\quad\bar\alpha\equiv
-{6\pi\over\bar\delta+i},\quad \bar\rho\equiv\rho/k^3,
\quad \bar{\sf G}'(\rv)\equiv {{\sf P}\cdot{\sf G}'(\rv)\over
i\kappa k^3}\,,
\eeq
where $\bar\alpha$ denotes the dimensionless atomic polarizability,
and the ansatz $\Pv_1(\rv)=P\,\pol\exp{(ik'z)}$ with Im$(k')>0$,
we obtain the susceptibility of the sample as
\beq
\chi={k'^2\over
k^2}-1={\bar\alpha\bar\rho\over1-\bar\alpha\bar\rho/3+C}\,,
\label{sus}
\eeq
with
\bea
C &=& -\bar\rho\int d^3\bar{r}\,\pol^*\cdot \left[{\bar\alpha^3
\bar{\sf G}'^3e^{-i\bar{z}}+ \bar\alpha^2\bar{\sf G}'^2\over
1-\bar\alpha^2 \bar{\sf G}'^2}\right]\cdot\pol\nonumber\\
&&-\bar\rho\int d^3\bar{r}\, \varphi \pol^*\cdot\left[
{\bar\alpha \bar{\sf G}' e^{-i\bar{z}}+
\bar\alpha^2\bar{\sf G}'^2\over 1-\bar\alpha^2\bar{\sf G}'^2}
\right] \cdot\pol\,.
\label{c}
\eea
Here we use the dimensionless integration variable $\bar\rv=k\rv$.
The quantity $C$, and hence also $k'$, have been forced to be independent
of position by essentially ignoring the effects of the surface of the
atomic sample~\cite{JAV99}.

The second term in the denominator in Eq.~{(\ref{sus})} gives the
Lorentz-Lorenz shift. In the absence of the $C$ term the electric
susceptibility is the standard column density result augmented with a
local-field correction. Atom statistics and atom-field collective effects
are encapsulated in the integral $C$. The expansion of Ref.~\cite{MOR95}
is such that the parameter $C$ takes into account
quantum statistical position correlations between any pair of atoms
and the exchange of photons between them to arbitrary order, but ignores
all repeated photon exchange involving more than two atoms.
Correspondingly, one may expand the functions inside the integrals in
Eq.~(\ref{c}) as power series in
$\bar\alpha \bar{\sf G}'$, and interpret the $n$th order as a
contribution in which a photon is radiated between a pair of atoms $n$ times.

In this paper we only consider the leading order in the modifications
of the optical response of an atom due to the presence of the other atoms,
and write
\beq
C\simeq -\bar\rho\bar\alpha\int d^3\bar{r}\,e^{-i\bar{z}}
\varphi(\bar\rv)\,\pol^*\cdot\bar{\sf G}'(\bar\rv) \cdot\pol\,.
\label{cexp}
\eeq
The expression~(\ref{cexp}) arises from processes in which any ``probe''
dipole is subject to the external driving field, and in addition to the
primary radiation from the other dipoles. Collective linewidths
and line shifts, processes that involve the repeated scattering of a 
photon between
the same atoms, are ignored. For uncorrelated locations of
the dipoles with $\varphi=0$,  the effects of the primary radiation from
the other dipoles on a  probe dipole average to zero. However, for the
FD statistics Eq.~{(\ref{cexp})} gives a nontrivial result. This reflects
the short-range ordering, within the correlation length, of the atoms in
the gas.

Fermions at $T=0$ fill the Fermi
sphere $\bar{n}_{\kv}=\Theta(k_F-|\kv|)$, with the Fermi wave number
$k_F=(6\pi^2\rho)^{1/3}$. In this case we may evaluate the pair
correlation function [Eq.~{(\ref{eq:DCF})}] in closed form. In the
thermodynamic limit the result is
\beq
\varphi(\rv)=-{9\over k_F^4 r^4}\[ {\sin k_Fr\over k_Fr}-\cos k_Fr\]^2\,.
\label{fervarphi}
\eeq
We also set the polarization of the incoming light field to be
in the $x$ direction. The propagator in Eq.~{(\ref{cexp})} then reads
\bea
\pol_x^*\cdot\bar{\sf G}'(\bar\rv)\cdot\pol_x &=&
{e^{i\bar{r}}\over4\pi}\left[ (1-\sin^2\theta\cos^2\phi)
{1\over\bar{r}}\right.\nonumber\\
&&\left.\mbox{}+ (3\sin^2\theta\cos^2\phi-1)({1\over
\bar{r}^3}-{i\over \bar{r}^2})\right]\,.
\label{proj}
\eea
After inserting Eqs.~{(\ref{fervarphi})} and (\ref{proj}) into
Eq.~{(\ref{cexp})} we obtain a (complicated) analytical expression, whose
density expansion reads
\beq
C={3i\over10}\( {\pi\over6}\)^{1/3}\bar\alpha
\bar\rho^{2/3}+{\cal O}(\bar\rho)\,.
\label{CFIN}
\eeq

In our 1D electrodynamics, the entire expression (\ref{c}) may be
integrated analytically for a FD gas at $T=0$~\cite{JAV99}. It is then
easy to see that the lowest-order density contribution is correctly
introduced by the expansion~(\ref{cexp}). This is also the case in three
dimensions, although the demonstration is more indirect. First, the terms
in Eq.~{(\ref{c})} that do not depend on
$\varphi(r)$ are linearly proportional to $\bar\rho$. Second, expanding
the  contributions to Eq.~{(\ref{c})} that {\em do\/} depend on $\varphi$
into a series of $\bar\alpha \bar{\sf G}'$, for orders beyond the one
included in~(\ref{cexp}) we find radial integrals of the form
\bea
&&\int d\bar{r}\, \bar{r}^2 {e^{\xi i\bar{r}}\over \bar{r}^{n-1}}
\varphi(\bar{r})
\propto {\cal O}(\bar\rho),\nonumber\\
&&\int d\bar{r}\, \bar{r}^2 {1\over \bar{r}^{n}}
\varphi(\bar{r})
\propto {\cal O}(\bar\rho)^{n/3}\,.\nonumber
\eea
Here $\xi$ are integers independent of $\bar\rho$, and $n\ge3$. Most
of these integrals formally diverge at the origin, but in a manner that
must eventually cancel to give a finite result~(\ref{c}). Besides, the
divergences do not depend on density. Omitting the divergences, the
integrals scale with density as indicated. All told, Eq.~(\ref{cexp}) not
only represents the lowest-order correction to the optical properties in
terms of the number of microscopic optical interaction processes  between
the atoms, but it also correctly gives the leading density correction
$\propto\bar\rho^{2/3}$ to the optical response.

\begin{figure}
\begin{center}
\leavevmode
\epsfig{
width=7.5cm,file=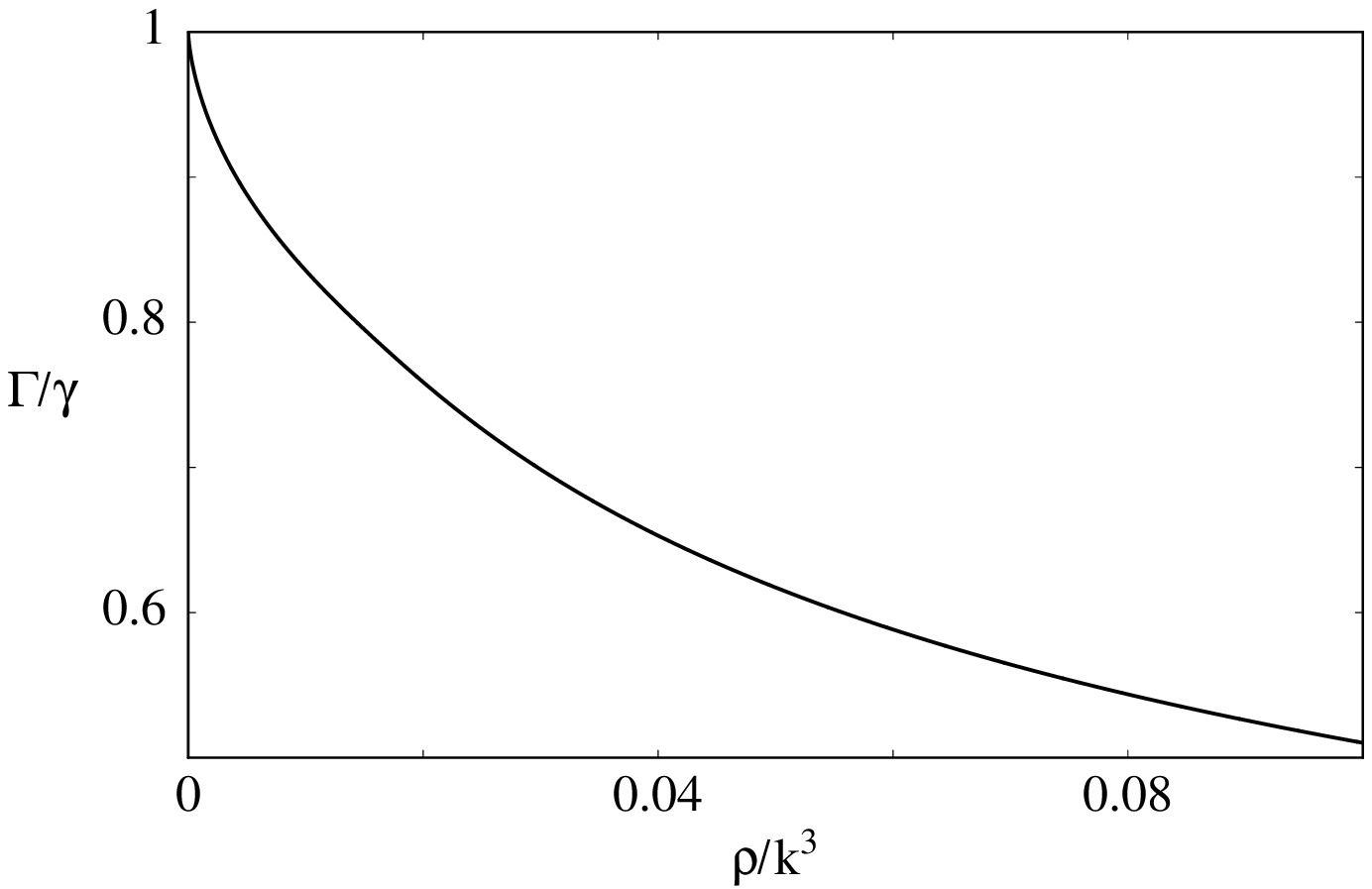}
\epsfig{
width=7.5cm,file=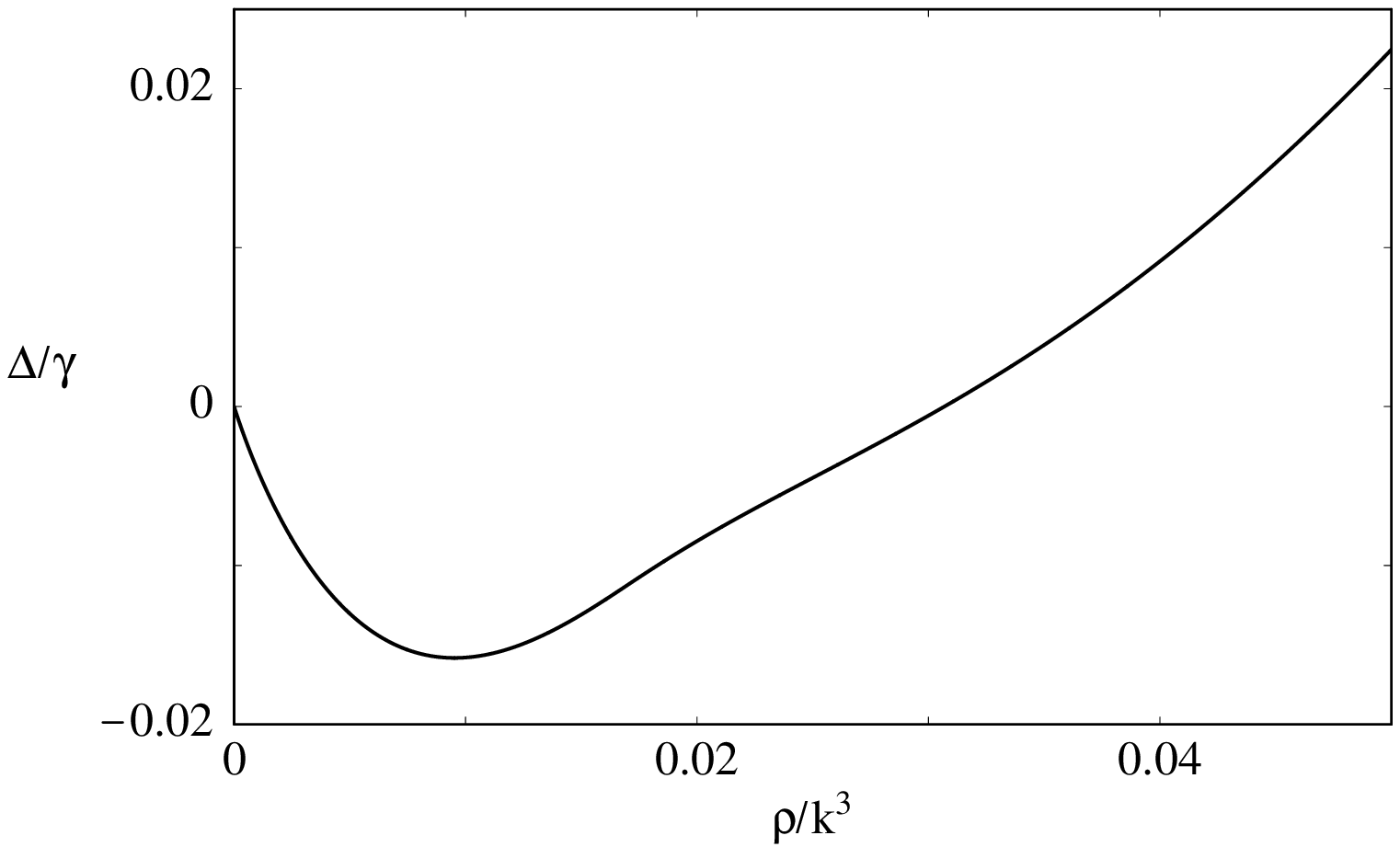}
\end{center}
\caption{The optical (a) linewidth and (b) the line shift of a
Fermi-Dirac gas as a function of the atomic density per cubic optical
wave number of the driving light $\bar\rho=\rho/k^3$. 
}   
\label{fig:1}   
\end{figure}

We have plotted the linewidth and the line shift, including both the
effect of the FD statistics and the Lorentz-Lorenz shift,
\bea
\Gamma &=& \gamma \left[1-6\pi{\rm Im}\left({C\over\bar\alpha}\right)
\right],\nonumber\\
\Delta &=&
\gamma\left[2\pi\bar\rho-6\pi{\rm Re}\left({C\over\bar\alpha}\right)
\right]\,,
\eea
as a function of density in Fig.~{\ref{fig:1}} using the full form of
$C$ from of Eq.~(\ref{cexp}).  A zero-temperature FD gas exhibits  a
striking linewidth narrowing already at low densities. For
$\bar\rho=1.5\times 10^{-3}$ the optical linewidth of the gas is
$\Gamma\simeq0.94\,\gamma$, and with $\bar\rho=1.5\times10^{-2}$ we have
$\Gamma\simeq0.79\,\gamma$. For the 767~nm transition in $^{40}$K
\cite{MAR98} the corresponding densities would be $\rho\simeq 8.2\times
10^{11}$ cm$^{-3}$ and $\rho\simeq8.2\times 10^{12}$ cm$^{-3}$. At
$\bar\rho=0.1$  ($\rho\simeq 5.5\times 10^{13}$ cm$^{-3}$) the optical
linewidth of a FD gas would be approximately half of the linewidth of an
isolated atom. However, at the latter density we may already have to
consider collective linewidths and line shifts to obtain a reliable
quantitative prediction, a task we do not undertake in the present paper.
From Fig.~\ref{fig:1}(b), the line shift is negative at low densities
and completely vanishes at $\bar\rho\simeq0.03$. At higher densities the
line shift turns positive.

The dramatic line narrowing may be attributed to the regular spacing
between the atoms characteristic of the FD statistics. As discussed in
Ref.~\cite{JAV99}, the mechanism  is particularly transparent in one
dimension. An alternative description of the line narrowing may be
obtained  in the momentum representation. At
$T=0$ the fermions fill the Fermi sphere. Due to the Pauli exclusion
principle only either strict forward scattering or scattering events that
take the recoiling atom out of the Fermi sea are allowed. The change of
the wave vector of an atom upon scattering satisfies
$|\Dkappav|=2k\sin(\theta/2)$, where $\theta$ is the scattering angle for
photons. All atoms are scattered out of the Fermi sea if
$|\Dkappav|>2 k_F$.  Thus, we see that for the photon scattering angles
$\theta$ satisfying
\beq
\sin(\theta/2)>k_F/k=(6\pi^2\bar\rho)^{1/3}\,,
\eeq
scattering is not inhibited by the FD statistics. On the other hand,
for $\sin(\theta/2)<k_F/k$ some recoil events would lead to an already
occupied state in the Fermi sea, and are forbidden. The suppression
of light scattering is strongest in the near-forward direction
corresponding to small values of $\theta$. When the density is
increased, at $\bar\rho\ge1/(6\pi^2)$ we have $k_F\ge k$, and
scattering is at least partially suppressed in all nonforward directions.
Correspondingly, Fig.~\ref{fig:1} shows that $\bar\rho=1/(6\pi^2)\simeq
0.017$ is a relevant scale for the density.

It is instructive to note the difference between different atom
statistics. For the Bose-Einstein condensate the standard factorization of
the correlation functions, $\rho_l=\rho^l$, corresponds to an uncorrelated
atomic sample, and gives $\varphi=0$. The leading correction to the 
standard column density linewidth therefore results from
the {\it cooperative} optical effects, the collective optical linewidths 
and line shifts, and it is proportional to atom density, just as
the Lorentz-Lorenz shift. FD statistics is different
because the correlations have a length scale $k_F^{-1}$
that enters the argument, {\em and} the length scale
itself depends on density: $k_F^{-1}\propto\bar\rho^{-1/3}$. The result
is that, at low densities, the effects of the FD statistics dominate over
the Lorentz-Lorenz shift. A Maxwell-Boltzmann ideal gas has
another nontrivial correlation function $\varphi$, but the length scale
is determined by temperature and does not depend on density. The leading
density correction to the optical response is then once more proportional
to atom density.

We have assumed a homogeneous FD gas in our analysis. Now, a FD gas may
be considered locally homogeneous~\cite{STO96} when the length scale over
which the density varies is much larger than the spatial correlation
length. Given the length scale of a harmonic trap
$l=(\hbar/m\omega)^{1/2}$ and the correlation length
$1/k_F=(6\pi^2\rho)^{-1/3}$ from Eq.~{(\ref{fervarphi})}, the criterion
reads $l^3\rho\gg 1$. A simple dimensional argument shows that this is
the same as requiring that the number of trapped atoms be much
larger than one. Furthermore, if the size scale of the atomic sample is
much larger than the wavelength of light, $l\gg\lambda$, it is
reasonable to expect that the refractive index as appropriate for the
local density applies in the bulk of the gas.

In conclusion, we have studied propagation of light in a FD gas. We have
discussed quantum statistical corrections to the refractive index, and
calculated the leading density correction to the standard column
density susceptibility. Already at low densities, fermions exhibit a
dramatic narrowing of the  resonance line. This might serve as a
signature of quantum degeneracy in a cold FD gas.

We acknowledge financial support from the European Commission through the
TMR Research Training Network ERBFMRXCT96-0066; from NASA, Grant No.\
NAG8-1428; and from NSF, Grant No.\ PHY-9801888.


\begin{references}
\bibitem{becexp} M. H. Anderson {\it et al.},  Science {\bf 269}, 198 (1995);
K. B. Davis {\it et al.}, Phys. Rev. Lett. {\bf  75}, 3969 (1995);
C. C. Bradley {\it et al.}, Phys. Rev. Lett. {\bf 75}, 1687 (1995);
D. J. Han {\it et al.}, Phys. Rev. A {\bf 57}, R4114 (1998);
U. Ernst {\it et al.}, Europhys. Lett. {\bf 41}, 1 (1998);
L. Vestergaard Hau {\it et al.}, Phys. Rev. A {\bf 58}, R54 (1998);
T. Esslinger {\it et al.}, Phys. Rev. A {\bf 58}, R2664 (1998); D. G.
Fried {\it et al.}, Phys. Rev. Lett. {\bf 81}, 3811 (1998).

\bibitem{JAV95b} J. Javanainen and J. Ruostekoski, Phys. Rev. A {\bf 52}, 3033
(1995).

\bibitem{STO96} H. T. C. Stoof {\it et al.}, Phys. Rev. Lett. {\bf 76},
10 (1996); M. Houbiers {\it et al.}, Phys Rev. A {\bf 56}, 4864 (1997).

\bibitem{BAR96} M. A. Baranov {\it et al.}, Zh. Eksp. Teor. Fiz. {\bf 64}, 273
(1996) [Sov. Phys. JETP Lett. {\bf 64}, 301 (1996)]; D. A. Butts and
D. S. Rokshar, Phys. Rev. A {\bf 55}, 4346 (1997); A. G. W. Modawi and
A. J. Leggett, J. Low. Temp. Phys. {\bf 109}, 625 (1998); G. M. Bruun
and K. Burnett,  Phys. Rev. A {\bf 58}, 2427 (1998); P. Stenius {\it et
al.}, preprint.

\bibitem{MAR98} B. DeMarco and D. S. Jin, Phys. Rev. A {\bf 58}, R4267
(1998).

\bibitem{JAV99} J. Javanainen, J. Ruostekoski, B. Vestergaard, and
M. R. Francis, Phys. Rev. A MS~AV6424; scheduled to appear in the January
1999 issue.

\bibitem{COH89}
C. Cohen-Tannoudji, J. Dupont-Roc, and G. Grynberg, {\it Photons and Atoms
} (Wiley, New York, 1989).

\bibitem{RUO97a} J. Ruostekoski and J. Javanainen, Phys. Rev. A {\bf 55},
513 (1997).

\bibitem{RUO97c} J. Ruostekoski and J. Javanainen, Phys. Rev. A {\bf 56},
2056 (1997).

\bibitem{JAC75} J. D. Jackson, {\it Classical Electrodynamics}, 2nd
ed. (Wiley, New York, 1975).

\bibitem{MOR95} O. Morice, Y. Castin, and J. Dalibard, Phys. Rev. A
{\bf 51}, 3896 (1995).



\end{references}
\end{document}